\documentclass[12pt]{iopart}

\usepackage{iopams} 
\usepackage{graphicx}
\usepackage{url}

\begin{document}

\title{A complete characterisation of the heralded noiseless amplification of photons}

\author{N. Bruno, V. Pini, A. Martin and R.~ T.~Thew}

\address{Group of Applied Physics, University of Geneva, Switzerland}
\ead{robert.thew@unige.ch}

\begin{abstract}

Heralded noiseless amplifcation of photons has recently been shown to provide a means to overcome losses in complex quantum communication tasks. In particular, to overcome transmission losses that could allow for the violation of a Bell inequality free from the detection loophole, for Device Independent Quantum Key Distribution (DI-QKD). Several implementations of a heralded photon amplifier have been proposed and the first proof of principle experiments realised. Here we present the first full characterisation of such a device to test its functional limits and potential for DI-QKD. This device is tested at telecom wavelengths and is shown to be capable of overcoming losses corresponding to a transmission through $20\, \rm km$ of single mode telecom fibre. We demonstrate heralded photon amplifier with a gain $>100$ and a heralding probability $>83 \% $, required by DI-QKD protocols that use the Clauser-Horne-Shimony-Holt (CHSH) inequality. The heralded photon amplifier clearly represents a key technology for the realisation of DI-QKD in the real world and over typical network distances.

\end{abstract}


\maketitle
\tableofcontents
\section{Introduction}

The concept of amplification in communication systems has long been used in the classical regime to overcome transmission loss. For quantum systems, amplification of quantum states is generally not possible due to the no (perfect) cloning theory - amplification will normally introduce noise, thus degrading the quality of the quantum state~\cite{Scarani2005}. However, heralded photon amplification can allow one to overcome transmission loss in a quantum channel, as it operates in a probabilistic fashion. Importantly, while it is probabilistic in nature, when successful, it provides a heralding signal that allows one to then perform subsequent operations. This heralding signal is what makes this approach interesting for Device Independent Quantum Key Distribution (DI-QKD) as it can herald the arrival of a photon (or qubit) and hence prepares the system so that the Bell test may be performed~\cite{Acin2006,Pironio2009a,Gisin2010}. More recently, it has also been incorporated into a quantum repeater protocol where it is used to herald the storage of a photon in a quantum memory~\cite{Mina2012}, opening the door to even greater distance for DI operations. It is clear that such a device could find wide spread use in myriad quantum systems where one needs to overcome inefficiencies associated with loss or multiple probabilistic operations, as well as where feed-forward signals can help in scaling complex quantum systems.

Noiseless photon amplification is related to quantum scissors~\cite{Pegg1998} and relies on quantum teleportation to herald the amplified state. It was first proposed by T.C. Ralph and A.P. Lund~\cite{T.C.Ralph2009} and has found several different implementations, either exploiting polarisation modes~\cite{T.C.Ralph2009,Xiang2010} , spatial modes in fibre optics~\cite{Osorio2012}, or using techniques such as single photon addition and subtraction~\cite{Fiur2009,ZavattaA.2011}. Also, it has shown potential application in discrete and continuous systems \cite{Fiur2012}.

In this article, we first present the principle operation of a heralded photon amplifier, then we introduce how our test device is realised.
The purpose of this article is to completely characterise the performance of the heralded photon amplifier at telecom wavelength, independently of the source and detector characteristics. Whe then discuss the operational limits of such devices and give some perspectives on further improvements in the context of DI-QKD.

\section{Principle of the heralded photon amplifier}

\begin{figure}[h!]
\includegraphics[width=\columnwidth]{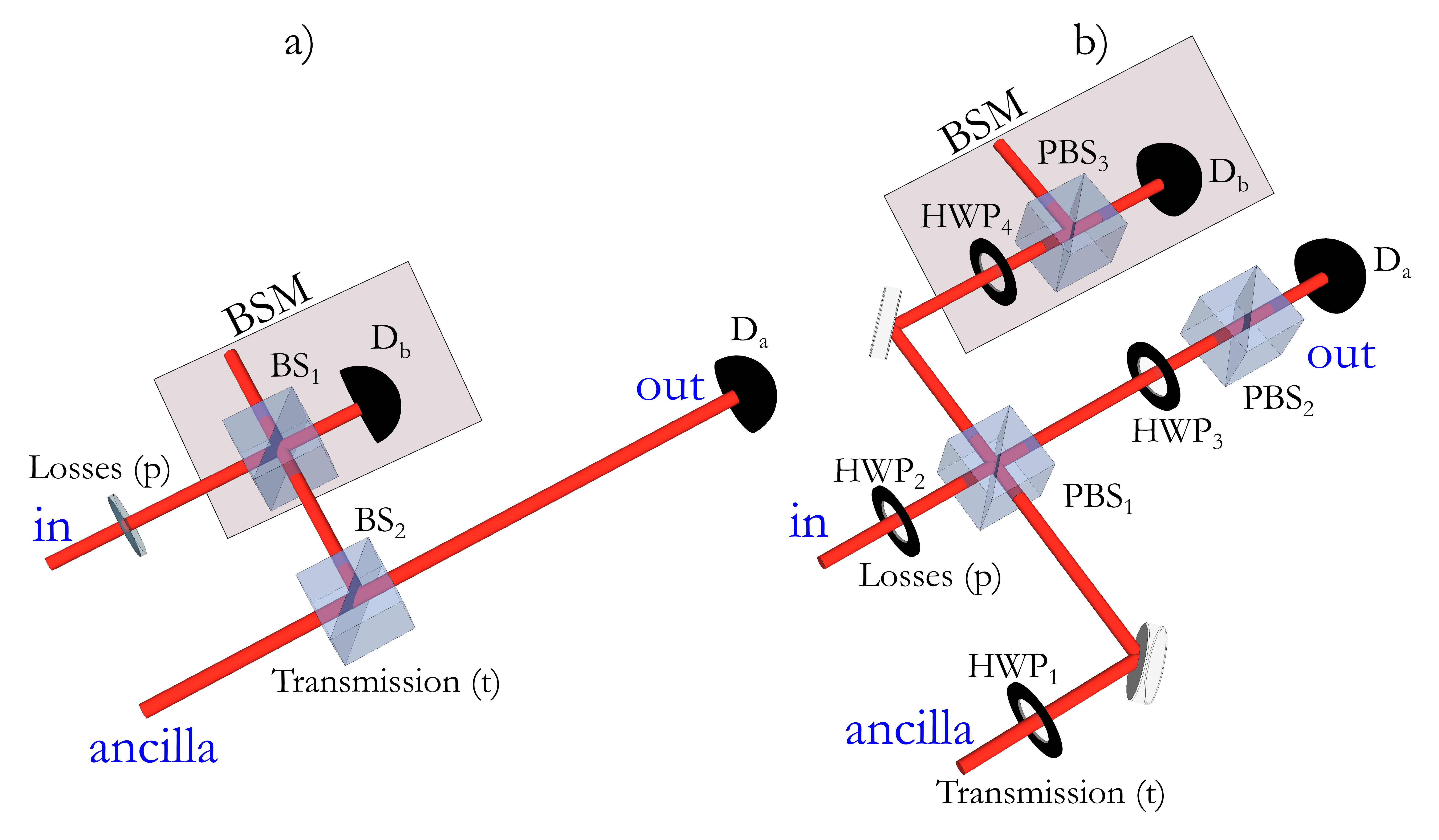}
\caption{$\mathsf{a})$ Standard representation of heralded photon amplifier, composed by a beam splitter with variable transmission $t$ and a balanced beam splitter followed by a photon detector. $\mathsf{b})$ Experimental setup for our test-device based on bulk optics and using, instead of spatial, polarisation modes (see main text for details).}
\label{setup}
\end{figure}

The concept of a heralded photon amplifier is illustrated in \figurename{~\ref{setup}.a}. The incoming state that we are interested in is usually a single photon that has been mixed with some vacuum due to transmission loss and has the form $ \rho_{in} = p\vert 0 \rangle \langle 0 \vert + (1-p) \vert 1 \rangle \langle 1 \vert$. An ancilla photon is first used to generate single photon entanglement~\cite{Sciarrino2002}. The input state is then combined on a beamsplitter with one mode of the entangled state and subsequently one photon is detected. This corresponds to a Bell state measurement. This requires that the two photons are indistinguishable and that the detector ($D_b$, see \figurename{~\ref{setup}.a}) can resolve the number of photons. If the initial ancilla state is maximally entangled, i.e. for a transmission of $t=0.5$, then this corresponds directly to the standard teleportation scenario. However, if we now vary this transmission we can bias the output state such that it has the form
\begin{equation}
\rho_{out} = \frac{1}{N(t)}\left[ p \vert 0 \rangle \langle 0 \vert + g^2(t) (1-p) \vert 1 \rangle \langle 1 \vert \right] 
\end{equation}
Where $g^2 ( t) = t/(1-t)$ is the gain factor, while $N(t) = p + g^2(t) (1-p) $ is a normalisation factor. 
The renormalised gain is defined as the ratio between the probability for the single photon component before and after the amplification and is given by \begin{equation}
G(t) = \frac{g^2(t)}{p + g^2(t) (1-p)} = \frac{t}{(1-t)p + t (1-p)}.
\label{gaineq2}
\end{equation}
We see that for $ G(t=\frac{1}{2}) = 1 $ the protocol reduces to a teleportation of the input state, and the gain is then greater than 1 for $ t > \frac{1}{2} $. 

One can notice that the gain depends on both $p$ and $t$. In particular, $G$ tends to infinity as $p \rightarrow 1$ (high losses) and $t \rightarrow 1$ (high transmission). However, a high gain doesn't imply a high heralding probability, which on the contrary is inversely proportional to the losses.

In addition, one should note that equation~(\ref{gaineq2}) doesn't take into account that, in practice, we have non-photon-number resolving detectors and non-zero losses through the components of the amplifier. This can result in a reduction in the actual experimentally achievable gain for a fixed input state and, in general, change the response of the amplifier as a function of $t$.

\section{Experiment}

In practice, our test device is of the form represented in \figurename{~\ref{setup}.b}, as this provides a more flexible setup for testing. The half wave plate and the polarising beam splitter ($\rm HWP_2 + PBS_1$) are used to simulate losses on the input state, while $\rm HWP_1$ and $\rm PBS_1$ play the role of a beam splitter with variable splitting ratio to define the transmission (t).
 
In this experiment, both the input and the ancillary photons belong to the same pair created in a type II spontaneous parametric down conversion process. For this purpose a $2\,\rm cm$ periodically poled Lithium-Niobate crystal is pumped by a mode locked Ti:Sapphire laser at $780\,\rm nm$, pulsed in the picosecond regime. 
The photons at 1560\,nm are filtered down to $1\,\rm nm$ by an interference filter to eliminate spectral distinguishability, before being separated by a polarising beam splitter. 
Coupling into single mode fibre with $\sim50 \% $ efficiency, ensures a well defined spatial mode.
Photon counting is performed by using two gated avalanche photo-diodes (APDs [IDQ-210]) with $25\%$ detection efficiency, $3\,\rm ns$ gate and a noise probability of $10^{-5}$ per gate that are synchronised with the laser. 
In all the performed measurements the laser is used to trigger one detector at 80 MHz, which, in turn, triggers the second one.

To ensure indistinguishability between the two photons in all degrees of freedom, a Hong Ou Mandel (HOM) type interference measurement is performed \cite{Hong1987}, using the two polarisation modes at $PBS_3$ \cite{Martin2010}. 
Following the setup reported in \figurename{~\ref{setup}.b}, it can be seen that in order for the amplification to take place the two photons are required to arrive at the same time on the $\rm PBS_1$. From this point they travel through the same optical path until they arrive at $HWP_4$ in two orthogonal polarisation states: $\vert H \rangle \vert V \rangle$. Here, the polarisation is rotated by $\pi /4 $ : $\left( \vert H \rangle + \vert V \rangle  \right) \left( \vert H \rangle -\vert V \rangle\right) $. If there is perfect indistinguishability the terms  $\vert H \rangle \vert V \rangle $ and $\vert V \rangle \vert H \rangle $ interfere and vanish, therefore two detectors at the outputs of $PBS_3$ will not click in coincidence. 
The measured net HOM visibility is $0.98\pm0.03$, with a pair creation probability per pulse $p = 0.01$. The visibility is limited only by double pair emission, and is in good agreement with the theory~\cite{Sekatski2012}, indicating that all degrees of freedom are well controlled in the experiment. 
\begin{figure}[h!]
\begin{minipage}[b]{0.49\columnwidth}
a)
\centering
\includegraphics[width=\columnwidth]{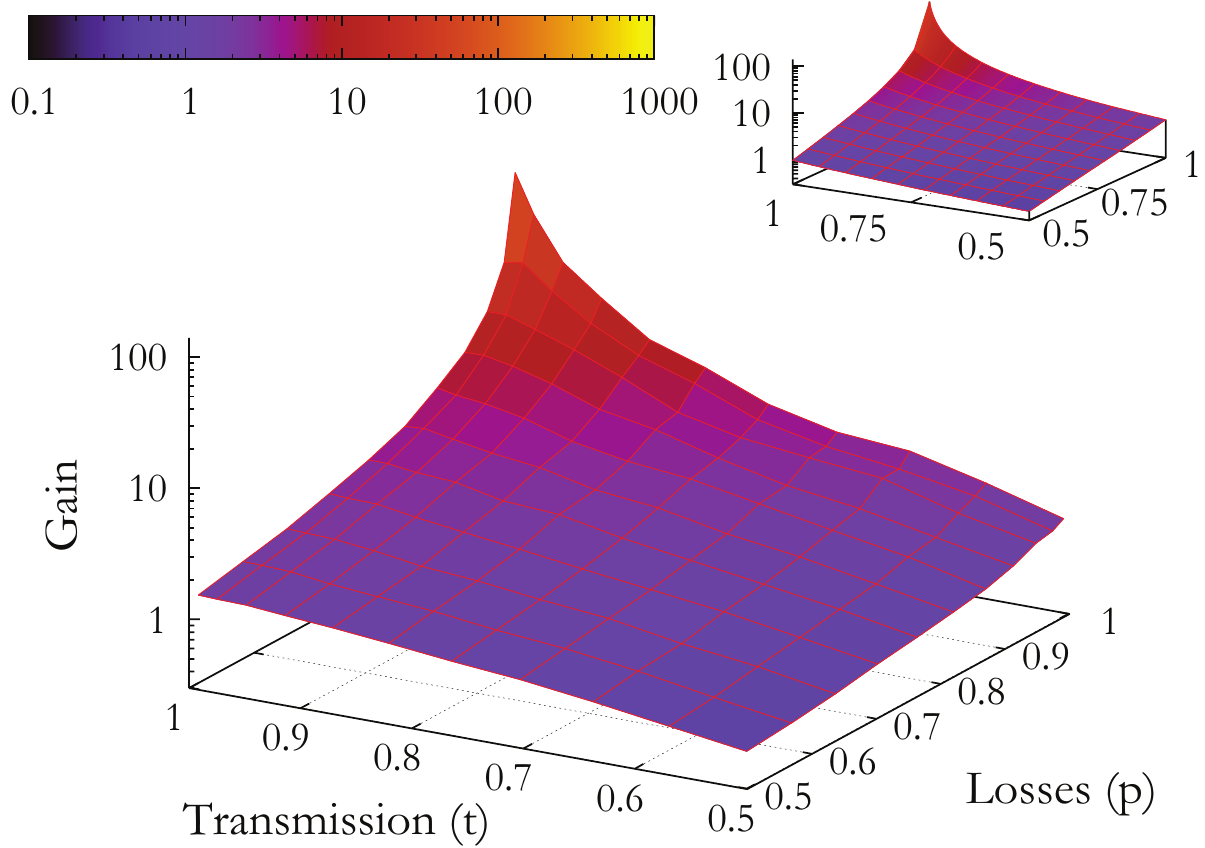}
\end{minipage}
\ \hspace{2mm} \hspace{3mm} \
\begin{minipage}[b]{0.45\columnwidth}
b)
\centering
\includegraphics[width=\columnwidth]{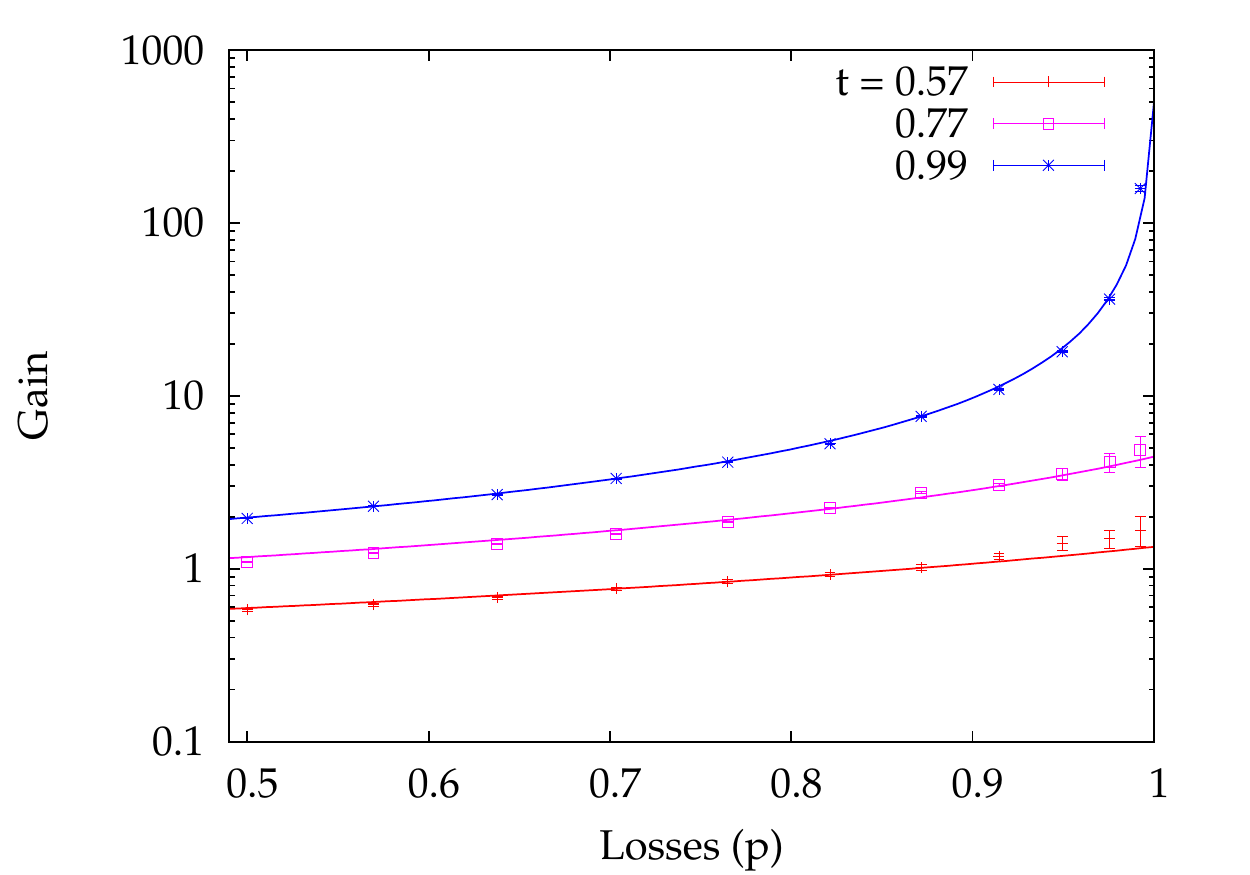}
\end{minipage}
\caption{a) Gain of the photon amplifier as a function of the losses introduced by rotating $\rm HWP_2$ and the transmission of the beam splitter tuned by $\rm HWP_1$. The result is in good agreement with the theoretical simulation (inset), which takes into account losses and non photon number resolving detection. b) Example of the gain as a function of the losses for three different transmissions.}
\label{gain}
\end{figure}

The gain is evaluated as the ratio between the probability of having a photon in the output state and the probability of having a photon in the input: $G = p_{out}/p_{in}$ .
We estimate the input probability $p_{in}$ as the ratio between the rate of triggers (counts in $\rm D_a$) and the rate of coincidences with $\rm D_b$. The input losses are varied between $0.5$ and $1$ by turning the waveplate $\rm HWP_2$.
In a second measurement the detector $\rm D_b$ triggers $\rm D_a$, and the output probability $p_{out}$ is given by the ratio between singles in $\rm D_b$ and coincidences with $\rm D_a$. For each value of input loss we vary the amplifier transmission $t$ between $0.5$ and $1$.

The gain $G$ is measured as a function of the transmission of the amplifier for eleven loss values. As shown in \figurename{~\ref{gain}}, the resulting gain is in agreement with the theoretical prediction taking in account losses, detection efficiency and the use of non-photon-number resolving detection. \figurename{~\ref{gain}.b} shows the curves for three fixed values of t as a function of p. We notice that the gain is measured to be $>100$ for the limit of high losses (p) and high transmission (see \figurename{~\ref{gain}.a}, blue stars), but in this regime the performance of the heralded photon amplifier tend to be less efficient in terms of success probability, as we will see in the following analysis. 

\begin{figure}[h!]
\begin{minipage}[b]{0.49\columnwidth}
a)
\centering
\includegraphics[width=\columnwidth]{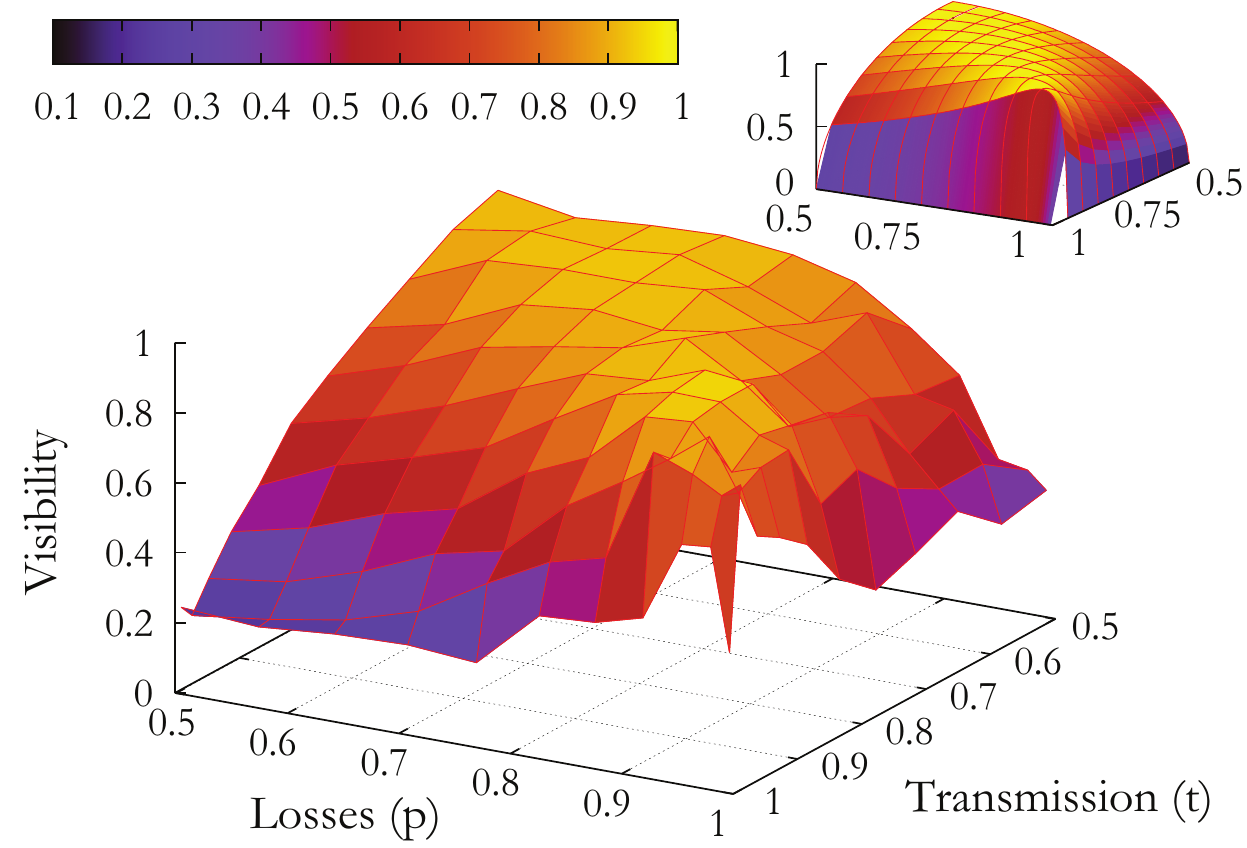}
\end{minipage}
\ \hspace{2mm} \hspace{3mm} \
\begin{minipage}[b]{0.45\columnwidth}
b)
\centering
\includegraphics[width=\columnwidth]{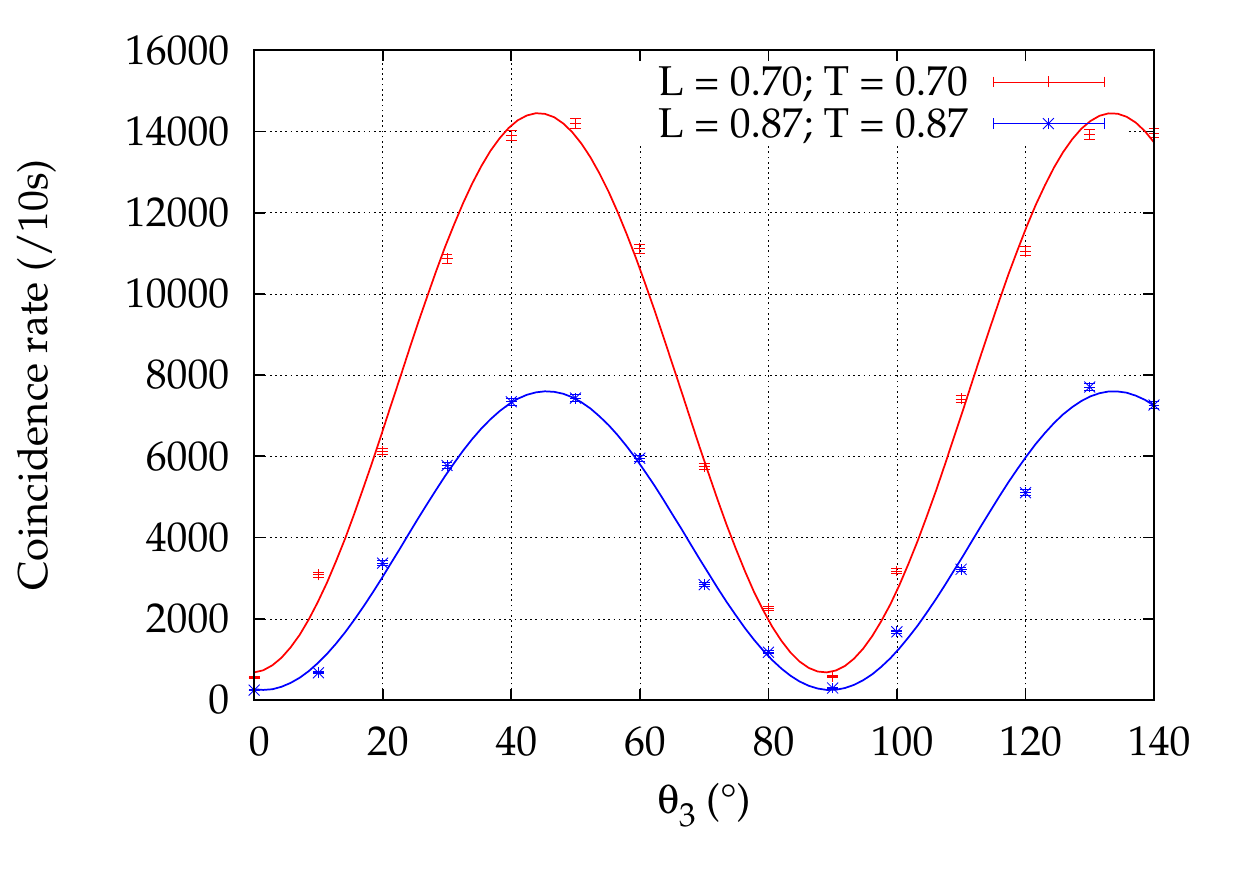}
\end{minipage}
\caption{\label{visi}a) Single photon interference visibility measured as a function of input losses and transmission of the amplifier, the theoretical prediction is represented in the inset. b) Interference fringes for two measurement settings such that the visibility is maximal, i.e. losses and transmission are balanced. The visibilities are $0.92 \pm 0.01$ and $0.94 \pm 0.03$ for the red (dots) and the blue (stars) curves, respectively. Visibilities are mainly limited by double pair emission in the down conversion process and polarisation dependent losses in optical elements in the setup, which change the weight of the two interfering modes.}
\end{figure}

The input state, as we already mentioned, is separated into two modes after $\rm PBS_1$, the ratio depending on the angle of $HWP_2$. The reflected and the transmitted modes correspond to the state to be amplified and the "lost" part, respectively. The latter is obviously not present in a communication channel, where lost photons are mainly absorbed or reflected, but in our case we can use it to look at the interference with the amplified mode.
Checking the coherence of the amplification process completes the characterisation of the device. The visibility is measured for each point represented in \figurename{~\ref{visi}.a} and found to be consistent with the expected behaviour. In particular, the visibility is maximal ($\sim 94 \%$) when $p$ and $t$ are complementary, i.e. the amplitudes of the two interfering modes are balanced. Changing the two parameters, the visibility inevitably decreases only because of imbalance in the amplitude, therefore it is still a proof of coherence. \figurename{~\ref{visi}.b} shows two examples of the measured interference fringes with maximal visibility.

\begin{figure}[h!]
\centering
\includegraphics[width=0.7\columnwidth]{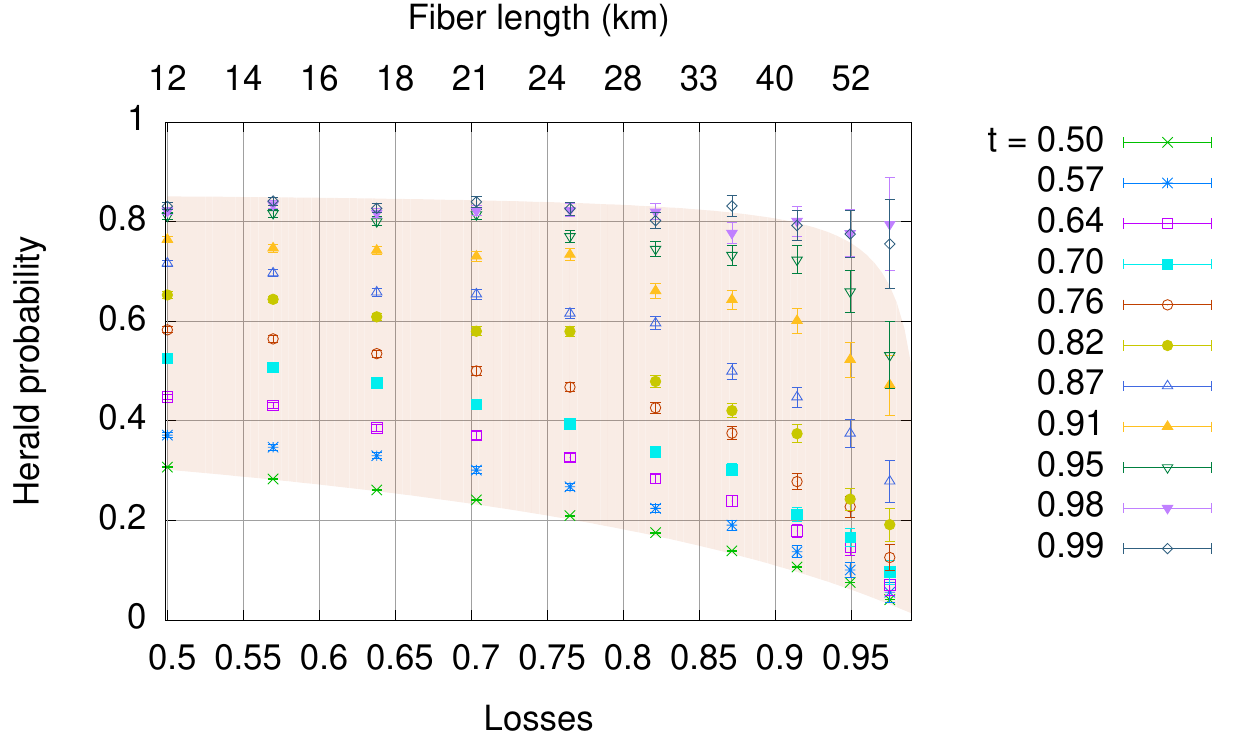}
\caption{Heralding probability as a function of the input losses, for the eleven settings of $t$. The probability is higher than $80\%$ when $t \geq 0.95$ even for losses of around $70\%$.
The colored region represents the theoretical prediction. The upper bound on the Heralding probability is given by the intrinsic losses of the amplifier.
On the upper axis, the equivalent transmission distance for an installed single mode fibre at telecommunication wavelength is given (losses are $0.24\, \rm dB.Km^{-1}$). In the extreme regime of high losses and high transmission the performance of the amplifier no longer follows the theory, because the trigger and coincidences rates fall to the noise level and the error bars significantly increase, as expected.}
\label{summary}
\end{figure}

Summarizing the result in a more intuitive way, as in \figurename{~\ref{summary}}, it's convenient to look at the heralding probability as a function of the losses introduced in the input state. With an amount of loss corresponding to the typical network distances, i.e. sending a photon through more than $ 20 \,\rm km $ of network installed fibre, it is still possible to have a heralding probability greater than $83 \%$. The results are renormalised taking into account the probability of pair emission and the losses before and after the device., i.e they consider the amplifier performances, and as such limited only by its intrinsic losses.

\section{Discussion and Conclusion}

We have fully characterised a heralded noiseless photon amplifier at telecom wavelength and obtained a gain $>100$ associated with a heralding probability greater than $83\%$ up to a distance in fiber of 20 km. Moreover, by duplicating the amplification stage it is possible to have a heralded polarisation qubit amplification~\cite{Kocsis2013}, which could allow the violation of a Bell inequality without the detection loophole by compensating for the losses~\cite{Gisin2010}. The heralded efficiency of such devices could be improved by reducing losses in the setup, in particular, by using anti-reflection coated and optimised optical elements~\cite{CunhaPereira2013}. However, for a more practical implementation of such a device, a fibre-based approach~\cite{Osorio2012} with a fixed gain for a fixed amount of loss would provide a realistically efficient solution with even lower internal losses, i.e. a heralding efficiency $>83\%$. One of the biggest challenges, though, is the generation of pure photons and coupling them into the heralded photon amplifier~\cite{CunhaPereira2013,Mosley2008,Evans2010,Rangarajan2011,Ramelow2013}. To resolve this problem, one could also think of a device completely realised with integrated optics on a chip, which includes a photon source and two variable couplers~\cite{Martin2012}, and potentially even detectors~\cite{Sprengers2011}.

\ack

We thank Nicolas Sangouard and Nicolas Gisin for useful discussions. This work was supported by the Swiss project NCCR-QSIT.

\section*{References}
\bibliographystyle{unsrt}
\bibliography{amplifier}

\end{document}